\documentclass[useAMS,usenatbib]{aa}  

\usepackage{graphicx}
\usepackage{txfonts}
\usepackage{graphicx} 
\usepackage{times}
\usepackage{epstopdf}
\usepackage{amsfonts}
\usepackage{amsmath}
\usepackage{amsbsy}
\usepackage{bm}
\usepackage{url}
\usepackage{microtype}
\usepackage{rotating}
\usepackage{sidecap}
\usepackage{longtable}
\usepackage{lscape}
\usepackage{rotating}
 \usepackage{array,multirow,graphicx}
\usepackage[export]{adjustbox}
\usepackage{amssymb,xcolor}
\usepackage{colortbl}
\usepackage{booktabs}
\usepackage{multirow}
\usepackage{url}
\usepackage{longtable}
\usepackage{arydshln}
\usepackage[table]{xcolor}
\usepackage[switch]{lineno}
\usepackage[utf8]{inputenc}

\begin{document} 


   \title{A MeerKAT view of the parsec-scale jets in the black-hole X-ray binary GRS 1758-258}

 \titlerunning{MeerKAT GRS 1758}
   \author{I. Mariani
          \inst{1}\inst{,2}
          \and
          S.E. Motta
          \inst{1}\inst{,3}
          \and
          P.Atri\inst{4}\inst{,5}
          \and
          J.H. Matthews\inst{3}
          \and
          R.P. Fender\inst{3}
          \and
          J. Martí\inst{6}
          \and
          P. L. Luque-Escamilla\inst{7}
          \and
          I. Heywood\inst{3}\inst{,8}\inst{,9}
          }

    \authorrunning{I. Mariani et al. }

   \institute{Istituto Nazionale di Astrofisica, Osservatorio Astronomico di Brera, via E.\,Bianchi 46, 23807 Merate (LC), Italy\\
              \email{isabella.mariani@inaf.it}
        \and
            Università degli Studi di Milano Bicocca, Dipartimento di Fisica, Piazza dell'Ateneo Nuovo, 1 - 20126, Milano Casella, Italy
        \and
        University of Oxford, Department of Physics, Astrophysics, Denys Wilkinson Building, Keble Road, OX1 3RH Oxford, United Kingdom
        \and
        ASTRON, Netherlands Institute for Radio Astronomy, Oude Hoogeveensedijk 4, 7991 PD Dwingeloo, The Netherlands
        \and
        Department of Astrophysics/IMAPP, Radboud University, P.O. Box 9010, 6500 GL Nijmegen, The Netherlands
        \and
            Departamento de Física, Escuela Politécnica Superior de Jaén, Universidad de Jaén, Campus Las Lagunillas s/n, 23071 Jaén, Spain
        \and
        Departamento de Ingeniería Mecánica y Minera, Escuela Politécnica Superior de Jaén Universidad de Jaén, Campus Las Lagunillas s/n, A3, 23071 Jaén, Spain.
        \and
        Department of Physics and Electronics, Rhodes University, P.O. Box 94, Makhanda 6140, South Africa
        \and
        South African Radio Astronomy Observatory, 2 Fir Street, Observatory 7925, South Africa
               }

   \date{Received month day, year; accepted xx xx, xx}

 
\abstract
{Jets from accreting black-hole (BH) X-ray binary systems are powerful outflows that release a large fraction of the accretion energy to the surrounding environment, providing a feedback mechanism that may alter the properties of the interstellar medium (ISM). Studying accretion processes alongside their feedback on the environment may enable to estimate the matter and energy input/output around accreting BHs.}
{We aim to study the extended jet structures around the BH X-ray binary GRS\,1758-258. First observed in VLA data, these parsec-scale jet structures originate from jet-ISM interaction, and are characterised by a peculiar Z-shape morphology.}
{Using the MeerKAT radio telescope we observed GRS\,1758-258 in L-band for a total exposure of 7 hr. Following a calorimetry-based method originally proposed for AGN and later applied to X-ray binaries, we estimated the  properties of the jets and of the surrounding ISM.}
{We detect a jet and a counter jet terminating in bow-shock structures induced by their interaction with the ISM. 
We identified both synchrotron and bremsstrahlung emitting regions within the northern lobe, while the southern lobe is dominated by thermal emission. We measured an ISM particle density between 10 and 40 cm$^{-3}$ across both the northern and southern jets, slightly lower in the northern region. The estimated ages of the two jet sides range from 6 to 51 kyr, with the northern jet seemingly younger than the southern one. The time averaged transferred jet-energy for both jets falls between $4.4\times 10^{33}$ and $3.3 \times 10^{36}$ erg s$^{-1}$, with slight differences between the northern and southern jets ascribed to different local environmental conditions. Comparing the new MeerKAT with archival VLA observations, we measured a proper motion of a portion of the northern jet of $\sim$ 130 mas/year.}
{Jet-ISM interaction structures on both sides of GRS\,1758-258 reveal different local ISM properties. The comparison between the morphology of this structures and those from other XRBs may indicate that the lobes in GRS\,1758-258 are younger and may result from a number of jet activity phases. 
The estimated time-averaged energy transferred to the environment is slightly lower than, but comparable to, that observed in other XRBs, consistent with the younger age of the lobes in GRS 1758-258 relative to those of other systems.
}


            
   \maketitle
%


\section{Introduction}

X-ray binaries (XRBs) are binary systems composed of a compact object such as a neutron star or a black hole (BH), and a companion star. XRBs  exhibit powerful emission triggered by accretion of matter from the companion to the compact object; this process often leads to the formation of fast, collimated outflows, called \textit{jets}. Jets are observed in systems across a wide range of mass scales in most accreting systems. Other than XRBs, they are observed in Active Galactic Nuclei (AGN), tidal disruption events and gamma ray bursts. Jets from XRBs and AGN are believed to release a large fraction of the accretion energy to the surrounding environment, providing a feedback mechanism that may alter the properties of the interstellar medium (ISM), or intergalactic medium (IGM) in the case of AGN. The consequences of such feedback include influencing the evolution of galaxies and enriching the chemical composition of the IGM in the case of AGN (\citealt{Magorrian1998}, \citealt{Croton2006}), as well as triggering/inhibiting nearby star formation and increasing or decreasing the density of the ISM through its energisation in the case of XRBs (\citealt{Heinz2008}, \citealt{Mirabel2015}).

The search for signatures of jet feedback from XRBs onto the ISM has been successful in a few XRB sources so far. The clearest cases are those of Cygnus\,X-1 (\citealt{Gallo2005}, \citealt{Atri_2025}) and GRS\,1915+105 (\citealt{Kaiser2004}, \citealt{Motta_2025}). Other examples include the the neutron star XRB Cir X-1 (\citealt{Sell2010}; \citealt{Coriat2019}) and the extragalactic BH XRBs in NGC 7793 (\citealt{Pakull2010}, \citealt{Soria2010}, \citealt{Hyde2017}), in the Large Magellanic Cloud \citep{Hyde2017} and in NGC 300 \citep{Urquhart2019}. Such signatures are reminiscent of the structures forming on much larger scales around radio-galaxies, where the effects of the jets interacting with the surrounding environment is clear (for instance in FR-I and FR-II sources, \citealt{Fanaroff1974}). 

The arched structure revealed by deep radio observations aligned with the resolved jets in Cygnus\,X-1 is the signature of the collimated jet-ISM impact (\citealt{Gallo2005}, \citealt{Atri_2025}). In GRS\,1915+105 an even more complex structure was identified tens of arcminutes away from the binary aligned to the direction of the jet: the radio spectral index indicates the presence of different emitting structures, characterized by synchrotron steep-spectrum emission and Bremsstrahlung flat-spectrum emission \citep{Motta_2025}. Both cases confirm the prediction of a radiative shock at the jet end, where the jet interacts with the ISM. This scenario is consistent with the model proposed by \cite{Kaiser2004}, an adaptation of a model originally proposed for AGN \citep{Kaiser1997}. The approach adopted by both \cite{Kaiser1997} and \cite{Kaiser2004} consists in using the jet-induced structures in the ISM (or IGM) as calorimeters to estimate the power of the jet launched by accreting systems that is transferred to the environment over the lifetime of the jet.

GRS\,1758-258 is a black hole XRB located at a distance of $\sim$ 8.5 kpc, towards the Galactic Center region. It exhibits bright and persistent X-ray emission \citep{Sunyaev_1991}, whereas at radio frequencies it shows a bright core and an two extended lobes at either side of 
it, a norther one and a southern one, generated by the approaching and receding jet, respectively. Both the core and the extended structures show a variability over timescales of a few years, discovered with the Very Large Array (VLA) during multiple observations conducted from 1992 to 2016 (see, e.g. \citealt{Marti2015}, \citealt{Marti2017nature}). The parsec-scale jets present a peculiar Z-shape structure, resembling the X- and Z-shape morphologies observed in \textit{Winged Radio Galaxies} \citep{Cheung_2007_wingedradiogalax}. 
These features and their temporal evolution are thought to originate from hydrodynamical instabilities in the jet, triggered by its interaction with the surrounding ISM. Numerical simulations of AGN jets show that the mechanical energy deposited at the jet head forms a bright spot. Subsequently, material from the jet inflates a \textit{cocoon}, as matter is funnelled sideways and backwards to create a backflow. The resulting low-density, high-pressure bright spot and cocoon drive a double-shock structure, with a bow shock propagating into the surrounding medium. A contact discontinuity separates the shocked jet material from the shocked ambient medium, although mixing can occur via the Kelvin–Helmholtz instability \citep[see, e.g., ][]{Matthews_2019}.

The apparent association with a CO cloud near GRS\,1758-258 identified in the survey of CO emission in the Galactic Plane (\citealt{Dame_COsurvey}, \citealt{Marti2017nature}) indicate  the presence of a gradient in the ISM density and supports the hydrodynamic scenario. \cite{Tetarenko_2020} observed GRS\,1758-258 using the ALMA interferometer, covering a frequency range of 84 -- 116 GHz. They reported the discovery of molecular emission lines $\sim$ co-spatial with the southern lobe continuum emission seen in the VLA observations: as the line detected is a density and shock tracer, the authors concluded that the observed structure confirms the jet-feedback scenario. However, no molecular emission was detected in the vicinity of the northern lobe. 

In this paper, we present the results of recent deep MeerKAT observations of GRS\,1758-258, which we also compared with previous VLA observations. The sensitivity achieved with MeerKAT allowed us to probe in detail the substructures within the jet-ISM interaction zones previously identified in VLA data. The paper is organized as follows: section 2 describes the observations and data reduction procedures. Section 3 summarizes the main results obtained from the data analysis. Section 4 outlines the model used to derive the jet calorimetry and presents the corresponding results. In section 5 we discuss our results and their implications, and in Section 6 we conclude the paper with our closing remarks.

\section{Observations and data reduction}
\label{Sec:Obs}

\subsection{MeerKAT data}
GRS\,1758-258 (J2000 18$^h$01$^m$12.48$^s$ -25$^{\circ}$44$'$35.69$''$) was observed three times with the MeerKAT interferometer as part of the BowKAT program (PI: Motta), which aims to study jet-driven bow shocks near Galactic BH-XRBs. The three observations were taken on April 4th, July 5th and August 15th, 2024, respectively. Data were acquired at L-band, at a central frequency of 1.284 GHz and a bandwidth of 0.85 GHz. The correlator was set up to provide 32768 channels with an 8-second mean integration time per visibility point. Each of the 3 observations had a duration of 3h 12m in total, with 2h 36m spent on target, 10m on the primary (flux and bandpass) calibrator J1939-6342 and 11m on the nearby secondary (phase) calibrator J1830-3602. The number of antennas used in the observations were, respectively, 60, 61 and 63 of the 64 available dishes, and the maximum baseline reached was 7.698 km. 

The observations were analyzed using OxKAT\footnote{\url{https://github.com/IanHeywood/oxkat}} \citep{Heywood2020}, a Python-based pipeline designed for the semi-automatic processing of MeerKAT data. After averaging the measurement sets down to 1024 channels, data from all fields (target and calibrators) were flagged using the CASA software \citep{CASAteam2022}. 
The first round of flagging consisted of removing the edges of the frequency bandpass, known bad frequency channels and zero amplitude visibilities. A second round targeted Radio Frequency Interference (RFI) in the time and frequency domains of the calibrators, prior to deriving the flux and bandpass calibrations from the primary calibrator and the delay and complex gain calibrations from the secondary. These correction were applied to the target field, whose data were then averaged in time (8s) and frequency (8 channels) and splitted to obtain a measurement set including only the target. Using the TRICOLOUR\footnote{\url{https://github.com/ska-sa/tricolour/}} package, specifically designed for RFI mitigation in MeerKAT data, another round of flagging was applied to the target measurement set. Finally, a first image of the $\sim 1.5 \ \text{deg}^2$
MeerKAT field was obtained using WSClean \citep{Offringa2012}. This image was used to generate an initial deconvolution mask, which was then applied to do a second round of imaging after self-calibrating the data using CUBICAL \citep{Kenyon2018} to solve for phase and delay corrections.

In order to remove potentially problematic artifacts from bright sources ($\sim$ 20 mJy and $\sim$ 30 mJy) spread across the target field, we decided to perform direction-dependent self calibration: this was done removing the two problem sources (RA = 18$^h$00$^m$31.49$^s$, DEC = -24$^{\circ}$47$'$21.29$''$ and RA = 18$^h$03$^m$42.58$^s$, DEC = -25$^{\circ}$44$'$02.19$''$) from the visibilities. After modelling the sources using WSClean, this time using 16 frequency channels, and once their model was partitioned into a FITS cube, we proceed to subtract them from the model visibilities of the entire field using CUBICAL. Finally, we obtain the \textit{peeled} image running WSClean on the processed data. 

\begin{figure*}[ht!]
\centering
\includegraphics[width=1\textwidth]{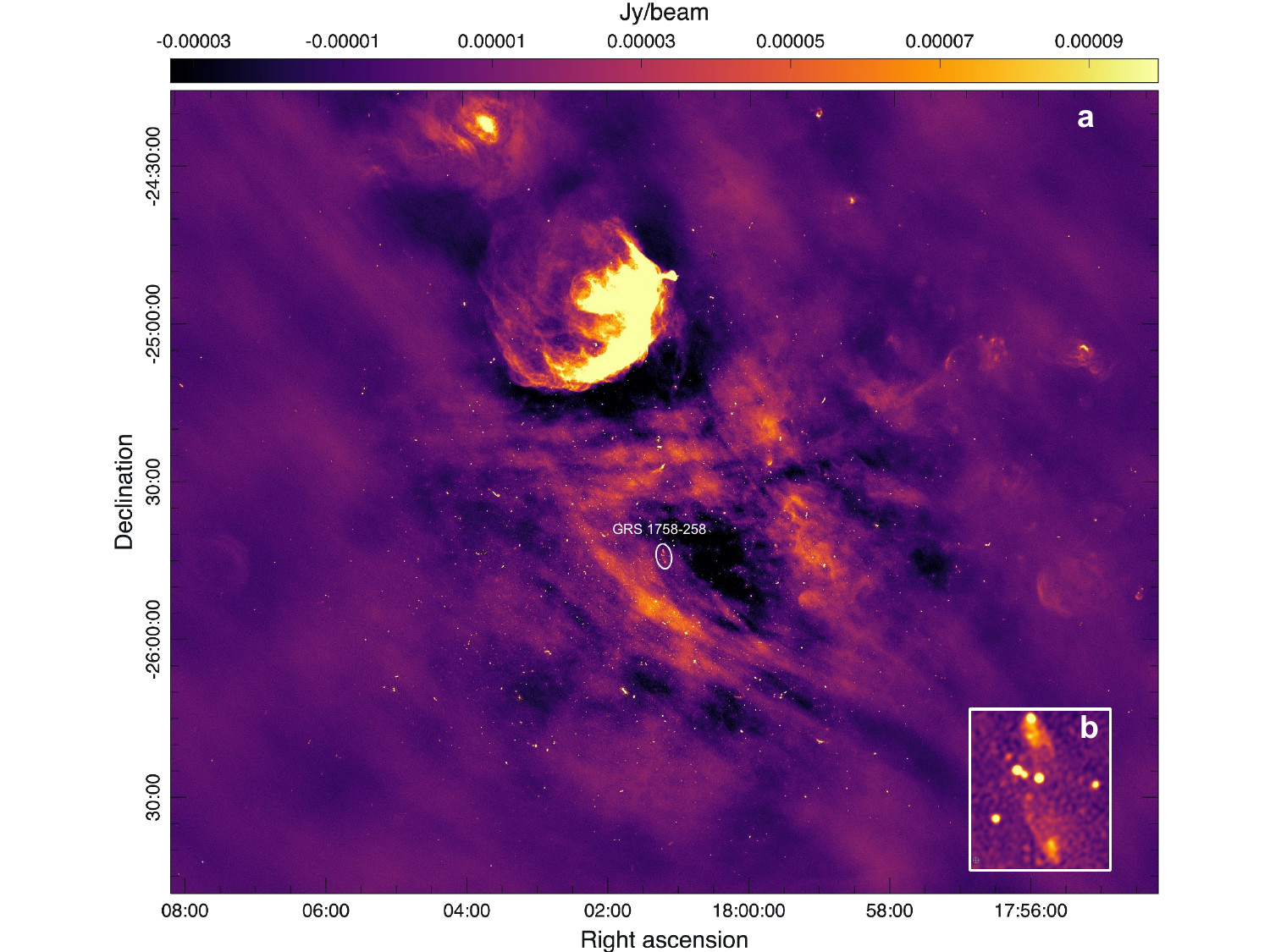}
\caption{\textbf{Panel a:} MeerKAT image of the sky field ($\sim 1.5$ deg$^2$) hosting GRS\,1758-258. The image was obtained from $\sim$ 7 hours of observation on target at 1.28 GHz. \textbf{Panel b:} Zoom-in view of the source GRS\,1758-258.}
\label{fig:TotalField}
\end{figure*}

The process described above was applied separately to the three measurement sets, meaning that the three observation blocks were calibrated and imaged individually. To maximize the signal to noise and enhance the detectability of the dim, extended structures around GRS\,1758-258, we used the task \texttt{concat} in CASA to combine the individual calibrated measurement sets into one single measurement set. The stacked measurement set, consisting of a total of 6h 48m spent on target, was self-calibrated again using CUBICAL and imaged with WSClean. The resulting radio map is shown in Fig. \ref{fig:TotalField}. 
For spectral analysis purposes, we performed the last run of imaging by also averaging the frequency band into 4 channels instead of 8, to maximize the sensitivity and the signal-to-noise ratio in each bin. The corresponding central frequency of the 4 channels were [0.963, 1.177, 1.391, 1.605] GHz. 

\subsection{VLA data}
We used the calibrated VLA data from \cite{Marti2015} and \cite{Marti2017nature}. For consistency with the MeerKAT data, we decided to image them using WSClean. To do so, we used the \texttt{importuvfits} task in CASA to convert the visibility files generated with AIPS\footnote{\url{http://www.aips.nrao.edu/index.shtml}} into a measurement set, required by WSClean. 

We used the observation from 1992, 1997, 2001, 2008 and 2016 in the C-band (4.8 GHz). We ran an initial round of WSClean to obtain a deconvolution mask for each epoch, which was then applied in a second round of imaging. We set the robust parameter to 1.0 for all the epochs except the 2016 data, where we used ROBUST = 2.0 to obtain beam sizes comparable to those of the MeerKAT data, namely 6.3$" \times$ 6.3 $"$. The beam size of all VLA archival images is $\sim$9.1$'' \times$ 4.2$''$, except for the 2001 data, where it is $\sim$12.0$'' \times$ 4.4$''$. The 2001 epoch was excluded from the subsequent analysis due to its limited signal-to-noise ratio, as explained in Section~\ref{sec:regionAB}.

\section{Results}\label{Sec:results}

We present the 1.5 deg$^2$ MeerKAT field in panel \textbf{a} of Fig. \ref{fig:TotalField} and a zoom-in of GRS\,1758-258 in panel \textbf{b}. A clear extended structure is detected around the position of GRS\,1758-258. The structure consists of a bright, compact core, and two elongated lobes, oriented at an angle of $\sim$ 9 deg with respect to the north-south direction on the plane of the sky. The northern and southern lobes correspond to the approaching and receding jets, respectively \citep{Marti2017nature}. Both lobes terminate in a bow-shaped structure, which is more extended to the right of the northern lobe and on the left of the southern lobe, respectively, forming a Z-shaped morphology. The apparent distances between the core and the northern and southern lobes are 1.20 arcmin and 1.57 arcmin, respectively, corresponding to 3.36 pc and 4.41 pc, assuming a source distance of 8.5 kpc and an inclination of 61$^{\circ}\pm$2$^{\circ}$ \citep{Bhuvana_2023_InclinationAngle} relative to the line of sight. The two extended structures forming the northern and southern lobes have been interpreted as evidence of the jets ejected from the binary system interacting with the surrounding ISM \citep{Marti2017nature}. 

\begin{figure}[ht!]
    \centering
    \includegraphics[width=\columnwidth]{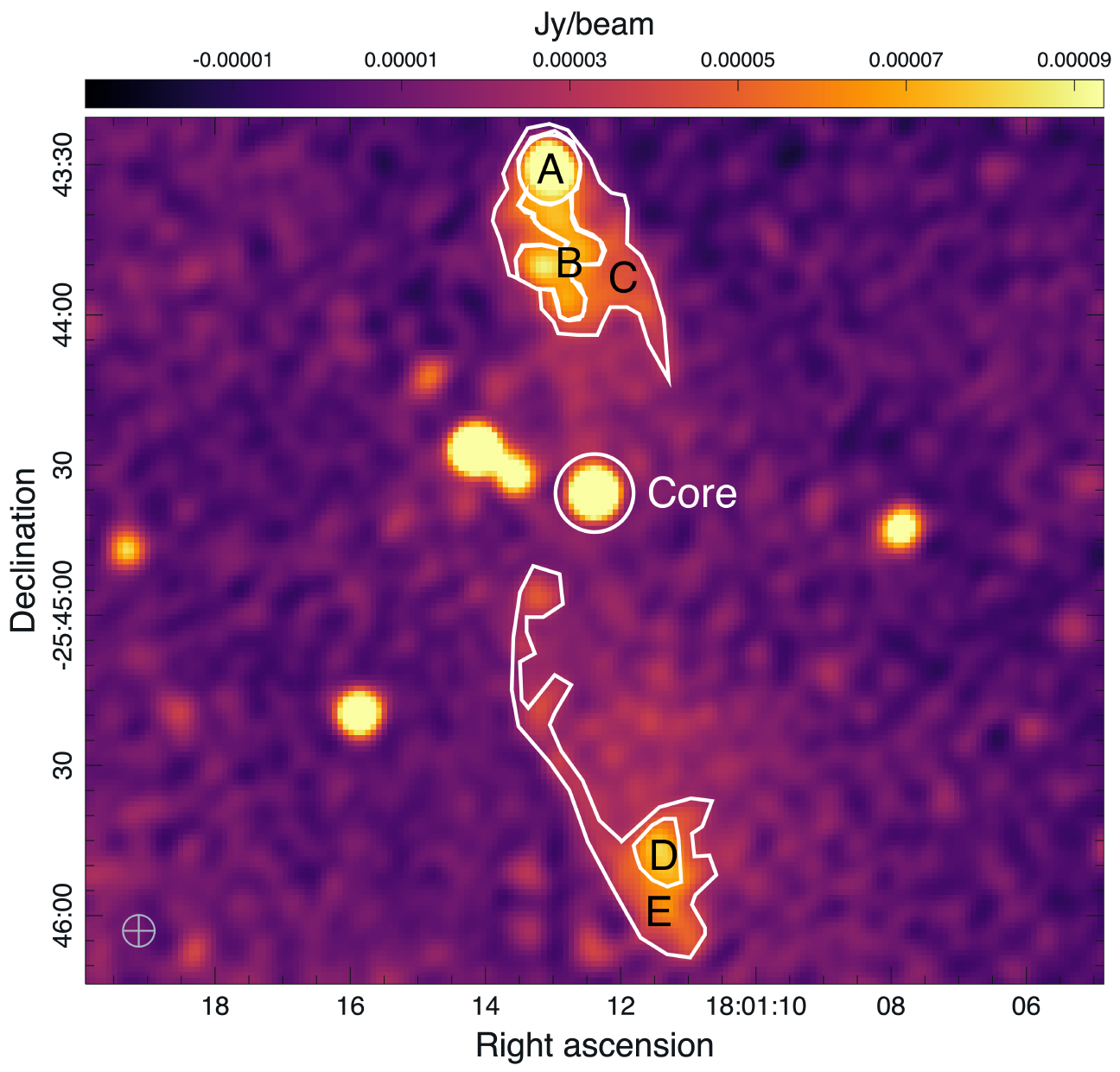}
    \caption{Zoomed-in view of GRS\,1758-258 and its extended lobes. The different regions used to model the jetted structures are outlined with white contours and labelled with letters from A to E.}
    \label{fig:ModelJets}
\end{figure}

\begin{figure*}[ht!]
    \centering
    \includegraphics[width=\textwidth]{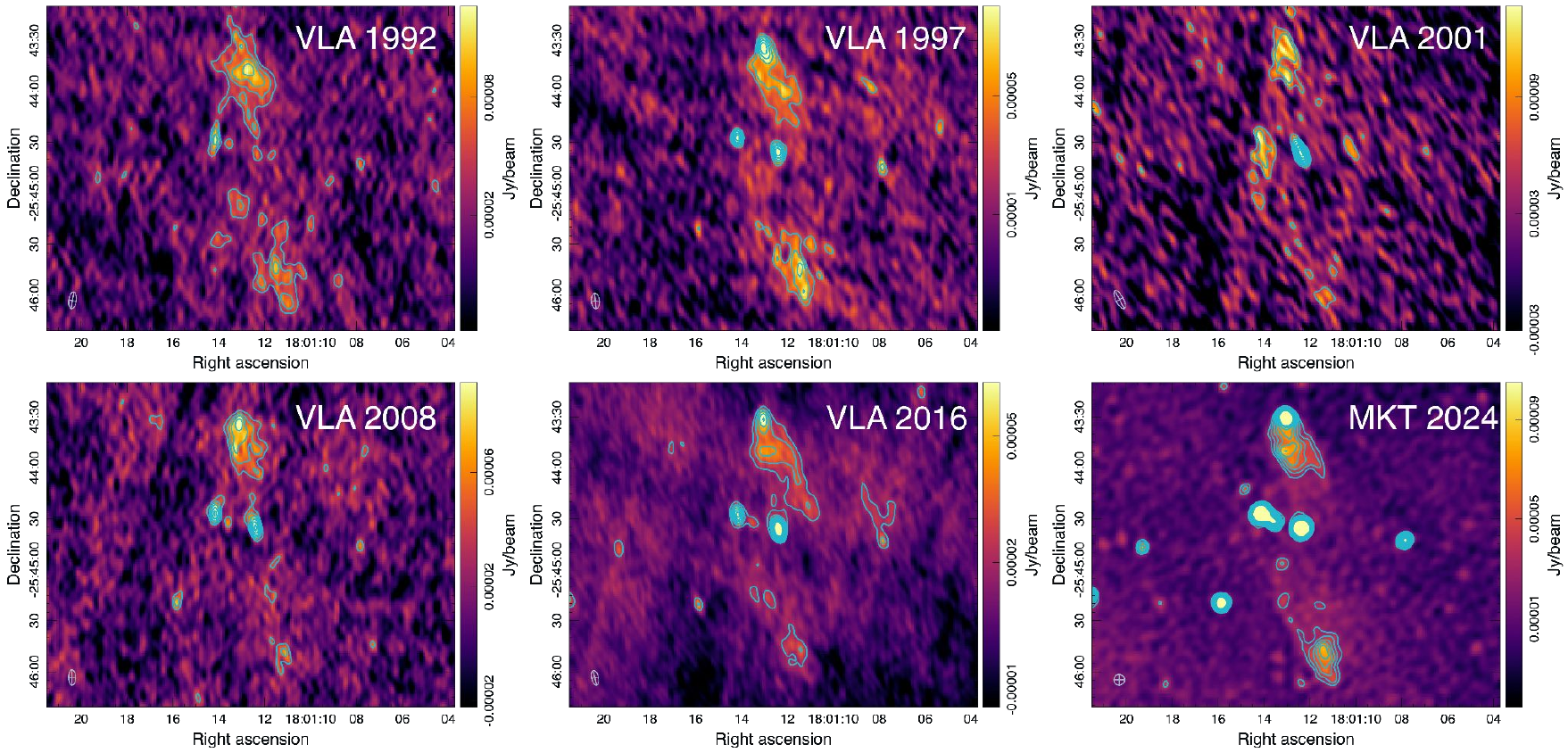}
    \caption{Comparison between the VLA archival observations and the new MeerKAT observations of GRS\,1758-258. Contours increase in eight steps, from 2 or 3 $\sigma$ from rms level depending on the epoch, up to 8$\sigma$. For the 2016 and 2024 epochs, which have a local rms of 10 $\mu$Jy, we adopted a threshold of 3$\sigma$. For the other epochs, which had higher local rms values (20 $\mu$Jy, except for the 2001 one, where it is 40 $\mu$Jy), a 2$\sigma$ threshold was used to ensure that the faint extended structures could still be identified.
    }
    \label{fig:Comparing_VLA-MK}
\end{figure*}

We measured the emission from the core using the tool \texttt{image fitting} in CARTA \citep{CARTA}, assuming a Gaussian model with FWHM corresponding to the beam width (6.3" $\times$ 6.3"). We obtained a flux of 0.424$\pm$0.018 mJy at RA = 18$^h$01$^m$12.400$^s\pm$0.004$^s$, DEC = -25$^{\circ}$44$'$36.10$''\pm4$0.06$''$. 
For the extended, resolved features corresponding to the northern and southern lobes, we modelled the emission defining the following regions, illustrated in Fig. \ref{fig:ModelJets}:
\begin{itemize}
    \item region A, also referred to the bright spot in the northern lobe;
    \item region B, also referred to the curved tail in the northern lobe;
    \item region C, representing the bow-shock structure\footnote{We define the \textit{bow-shock structure} as the feature resulting from a bow-shock induced by the interaction of the binary’s jet with the ISM.} 
    at the end of the northern jet-ISM interaction region;
    \item region D, a brighter spot identified in the southern lobe;
    \item region E, representing the bow-shock structure at the end of the southern jet-ISM interaction region.
\end{itemize}
We defined these regions considering structures with surface brightness at least 3$\sigma$ above the local average rms. The inner boundaries of the outer regions coincide with the outer boundaries of the inner regions, so that each region does not overlap with any other region.
The integrated flux of each region, compared with the values reported in \cite{Marti2002} from VLA data, are reported in Table \ref{tab:spect_index}. 
In Fig. \ref{fig:Comparing_VLA-MK} a comparison between the structure observed with MeerKAT and the one observed with the VLA (\citealt{martí2019VLAimage}, using data at 6 cm and 3.5 cm) is shown. 
The overall structures in the images are remarkably similar.

\begin{figure}[ht!]
    \centering
    \includegraphics[width=\columnwidth]{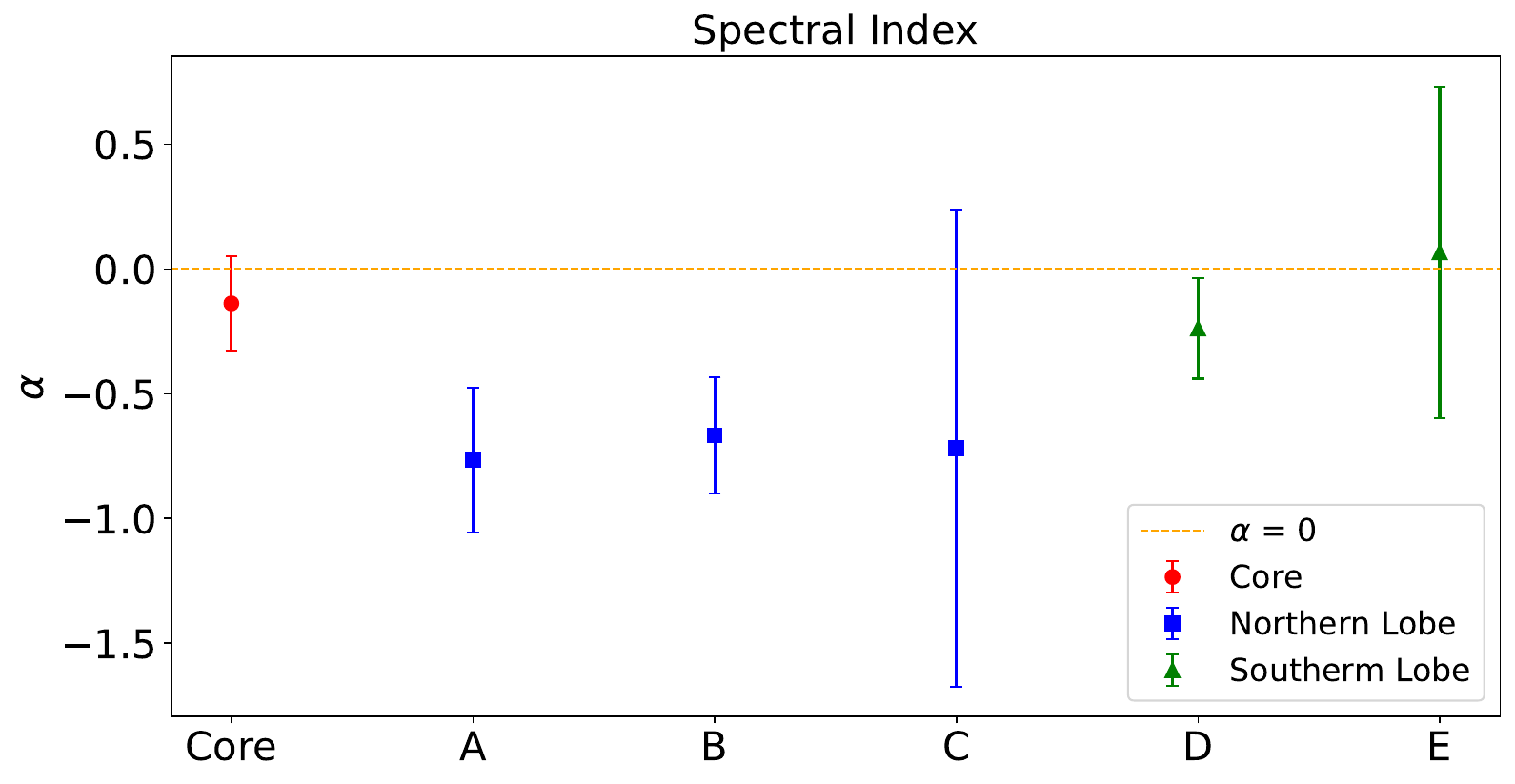}
    \caption{In-band spectral index of GRS\,1758-258 core and northern and southern lobe regions.}
    \label{fig:alpha}
\end{figure}

We measured the in-band spectral index for all the regions marked in Fig. \ref{fig:ModelJets} using the MeerKAT images obtained by stacking our observations and imaging in 4 spectral channels. We determined the spectral slope of the core emission measuring the fluxes in each individual channel using the same methodology described above.  
We fitted the resulting fluxes with a power law $S_{\nu} \propto \nu^{\alpha}$, where $S_{\nu}$ is the specific emission at the central frequency of each channel. The first channel was discarded due to the low signal-to-noise ratio. We used the same approach to estimate the spectral index of the extended regions A, B, C, D and E, using the integrated flux measured in each of the 3 selected channels (reported in Table \ref{tab:flux_spectra}). The results are illustrated in Table \ref{tab:spect_index} and Fig. \ref{fig:alpha}.  
The core and the region D and E in the southern lobe are consistent within uncertainties with a flat radio spectrum, whereas region A and B in the northern lobe exhibit a steep spectrum. The nominal spectral index value for region C is consistent with a steep spectrum, although considering its uncertainty a flat spectrum cannot be ruled-out. 
Our results are consistent with those from \cite{Marti2002}, where the authors considered the whole emission from the northern and southern lobes. 
Within uncertainties these results are also consistent with those reported by \cite{Hardcastle_2005}, who used additional VLA L-band observations to determine the spectral indices of the core and the two lobes.

\section{Modelling}\label{Sec:modelling}

To infer the properties of the emission observed along the direction of the two 
jets launched by GRS\,1758-258, we adopted the model described in \cite{Kaiser2004}, following the same approach of \cite{Motta_2025} and \cite{Atri_2025}. This model assumes that supersonic jets from XRBs can terminate in strong shock fronts when they interact with the ambient medium. The jets release energy and inflate two lobes that expand while remaining confined, being overpressured compared to the unshocked ambient medium. As they expand forward and sideways, the lobes generate a bow-shock structure.

GRS\,1758-258 exhibits a rather complex structure: the sensitivity achieved with the MeerKAT interferometer enables the identification of distinct emission regions within both the northern and southern jets. In the following sections we describe how we estimate the energy transport rate of the jet launched from GRS\,1758-258. Taking into account the significant uncertainties in some of the parameters used and the assumption made, we provide ranges of the quantity we estimated. These ranges are derived using a Markov Chain Monte Carlo approach to derive posterior distributions for each quantity. Unless otherwise specified, throughout this work we provide ranges corresponding to the 16th and 84th percentile.

\begin{table*}[ht!]
    \centering
    \begin{tabular}{cccccccc}
    \toprule
      & MeerKAT & $S_{1.28 \text{GHz}}$ [mJy] & $\alpha$ & VLA & $S_{\text{6cm}}$ [mJy] & $S_{\text{3.5cm}}$ [mJy] & $\alpha$ \\
    \toprule
    Core & &0.42$\pm$0.02 &-0.14$\pm$0.19 & C$^*$ & 0.14$\pm$0.02 & 0.15$\pm$0.02 &0.1$\pm$0.4  \\
    \midrule
       & A &0.31$\pm$0.03 & -0.77$\pm$0.29 & & & \\
     Northern Lobe & B &0.33$\pm$0.03 & -0.67$\pm$0.23 & B$^*$ &0.65$\pm$0.06 &0.40$\pm$0.05 &-0.9$\pm$0.3 \\
       & C &0.47$\pm$0.05 & -0.72$\pm$0.96 & & \\
    \midrule
       & D & 0.14$\pm$0.01 & -0.24$\pm$0.20 & &  \\
     Southern Lobe &   & & & A$^*$ &0.25$\pm$0.04 &0.24$\pm$0.03 & -0.1$\pm$0.4 \\
       & E &0.57$\pm$0.06 & 0.07$\pm$0.66 & & \\
    \bottomrule
    \end{tabular}
    \caption{Fluxes and spectral indices obtained using MeerKAT data compared to the ones from \cite{Marti2002} with VLA. The VLA regions are labeled following the convention of \cite{Marti2002} and are marked with an asterisk to distinguish them from our own nomenclature. With their higher sensitivity, MeerKAT data allowed to set a more detailed model of the lobes' regions; nevertheless the spectral slope values are consistent within their uncertainties.}
    \label{tab:spect_index}
\end{table*}

\subsection{Northern lobe}
\label{subsec4:noth}

Based on the radio map we divided the northern lobe in three regions, showed in Fig. \ref{fig:ModelJets}. The bright hotspot, region A, consists of an approximatively circular region, possibly marginally resolved. For simplicity, hereafter we treat region A as a point source. As mentioned above, we labelled the arched structure near region A as region C, and the tail located behind region A and extending towards the core as region B. 

To estimate the density of the shocked medium $\rho$ we need to know the type of the emission triggered in the jet-ISM impact region. Based on the spectral index results, we interpreted the emission in regions A and B as synchrotron radiation from active electrons within the jet, while for region C we assumed bremsstrahlung radiation, produced as the region cools down after being energized by a past jet interaction.
It is also possible that both synchrotron and bremsstrahlung emission contribute to the observed emission in these two regions, perhaps due to the effects of different ejecta launched at different times. The large uncertainties prevent us from determining whether a minor contribution from the other process is present in each region.

\subsubsection{Region A and B}
\label{sec:regionAB}

Since we considered both A and B as synchrotron emitting regions, we estimated their energetics assuming the minimum energy scenario, which occurs at equipartition. Following \cite{Longair1994}, we estimated the magnetic field at equipartition measuring the integrated luminosity and the volume of the two regions. We calculated the radio luminosity using the monochromatic fluxes listed in Table \ref{tab:spect_index}, integrating over a frequency range of $\approx 10^7-10^{11}$ Hz, which is a typical energy range for synchrotron emission.  

We modelled region A as a sphere with an apparent diameter equal to the beam size since the structure may be marginally resolved.
We assumed a filling factor $f=0.1$, which accounts for the volume effectively occupied by the gas. This is a reasonable assumption, considering the different spectral index values found within the northern lobe region, which suggest the presence of multi-phase gas, as discussed later in section \ref{sec5:morph_large_scale}. The use of a filling factor also mitigates possible uncertainties in the volume estimate.

The estimate of region B volume is trickier due to its irregular shape: for simplicity, we approximated it as a conical region, with the base set to the size of the beam and length $\approx$ 24 arcsec. For both regions, we found a luminosity of $\sim 10^{30}$ erg s$^{-1}$ and a volume of $\sim 10^{54}$ cm$^3$ (see Table \ref{tab:total_table_south} for the detailed values) 

We found a lower limit on the magnetic field of $B_{eq} \gtrsim 4.5 \times 10^{-5}$ G for both regions. Assuming equipartition, one can also estimate the corresponding total pressure inside the regions and the electron density contributing to the synchrotron emission: in our case, we obtained $p_{B_{eq}} \gtrsim 6.5 \times 10^{-11}$ erg cm$^{-3}$ and $n_e \lesssim 7 \times 10^{-7}$ cm$^{-3}$ (see Table \ref{tab:regAB} for the specific values and section \ref{App:synch} for the formulae used to compute these values).

The integrated luminosity and the magnetic field corresponding to the minimum energy we obtained are comparable to those of \cite{Marti2015} for the region A, which these authors identified as the northern hotspot. However, we obtained a significantly lower value for the particle density estimate ($\sim 10^{-6}$ cm$^{-3}$ in our case vs. $\sim 10^{-3}$ cm$^{-3}$). This difference is due to a different choice of synchrotron frequencies, much narrower in this work than in \cite{Marti2015} ($10^{7}-10^{11}$ Hz vs. $10^{2}-10^{15}$ Hz).

\begin{table}[ht!]
    \centering
    \begin{tabular}{ccc}
    \toprule
       & region A & region B \\
       \midrule
       $B_{eq}$ [G] & $\gtrsim 4.5 \times 10^{-5}$ & $\gtrsim 4.8 \times 10^{-5}$ \\
       $p_{B_{eq}}$ [erg cm$^{-3}$] & $\gtrsim 6.5 \times 10^{-11}$ & $\gtrsim 7.2 \times 10^{-11}$ \\
       $n_e^{\text{syn}}$ [cm$^{-3}$] & $\lesssim 6.9 \times 10^{-7}$ & $\lesssim 5.8 \times 10^{-7}$ \\
    \bottomrule
    \end{tabular}
    \caption{Magnetic field, pressure and particle density of the synchrotron-emitting regions, obtained under the assumption of equipartition.}
    \label{tab:regAB}
\end{table}

Comparing the images from VLA and MeerKAT observations we found that region A exhibits proper motion. 
As previously noted by \cite{Marti2017nature} the core does not exhibit any detectable proper motion, and the core position across different epochs remains consistent with each other within uncertainties. However, region A shows a clear motion over the years: we plotted the positions of both the core and region A in Fig. \ref{figA:regA-core-pos} and reported their positions in Table \ref{tab:regA_pos}. The positional errors were estimated following the methodology of \cite{Condon97}. We excluded the positions from epoch 2001 from the fit due to its low signal-to-noise ratio, which prevented us to clearly identify the position of region A. Using a simple linear function, we derived a proper motion of 100-160 mas years$^{-1}$, corresponding to a de-projected velocity of $\dot{L}_{\text{regionA}} \sim 4600-7500$ km s$^{-1}$ ($\sim$ 0.015 -- 0.025 c) and a Lorentz factor of $\gamma_{\text{jet}} \sim 1.0001 - 1.0003$. We note that the deviations from a linear trend of the (variable) positions of region A over time may be consistent with a precessing jet, as already proposed by \cite{L-Escamilla_2020}. Unfortunately, the quality of our data is insufficient to draw any firm conclusion in this context.

\begin{figure}[ht!]
    \centering
    \includegraphics[width=\columnwidth]{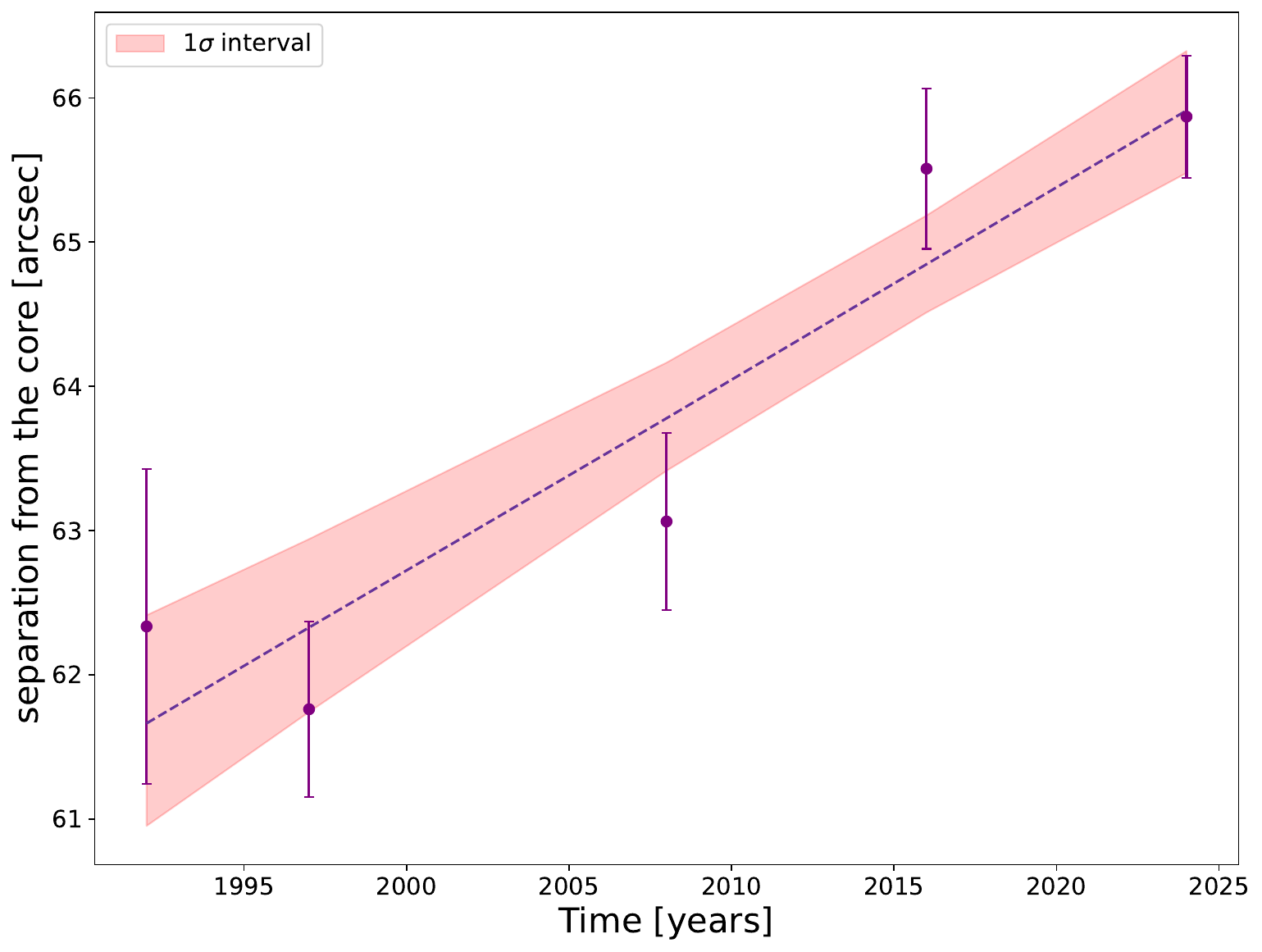}
    \caption{Proper motion of region A, measured by tracking its separation from the core across different epochs.}
    \label{fig:regA_separation}
\end{figure}

\subsubsection{Region C}
\label{sec:regionC}

Under the assumption of bremsstrahlung emission, we hypothesize a fully ionized hydrogen structure as bremsstrahlung radiation would not be efficient otherwise. The electron density of ionized hydrogen emitting bremsstrahlung depends the monochromatic emissivity (monochromatic luminosity per unit volume) and the gas temperature \citep{Longair1994}.

We considered a temperature range of T $\sim 10^4 - 10^6 \ \text{K}$: for T $\lesssim 10^4 \ \text{K}$ the hydrogen would not be fully ionized, while for T $\gtrsim 10^6 \ \text{K}$ the bremsstrahlung cooling becomes significant for particle densities typical for the ISM \citep{Gould1980}. This implies a jet velocity of $\dot{L}_{\text{jet}} \approx$ 21 -- 360 km s$^{-1}$ (assuming a mono-atomic shocked gas). Given the complexity of this region and the difficulty in measuring its volume, we calculated the emissivity of a cylindrical unit volume with length equal to the average thickness of the structure, and diameter equal to the beam diameter, emitting flux equal to the average flux measured for the entire region. We obtained a shocked particle density in the range of $\sim$ 55 -- 95 cm$^{-3}$ (eq. \ref{eq:ne} in App. \ref{App:App1}), which corresponds to a pre-shock density of $\sim$ 10 -- 25 cm$^{-3}$, as we are assuming that the shocked medium is four times denser than the un-shocked medium. This values are consistent with the typical HII and ISM regions electron densities, i.e. 0.1 -- 100 cm$^{-3}$.

\subsection{Southern lobe}

Similarly to what we did for the northern lobe, we modelled the structure of the southern lobe by defining two regions: D is the brighter and more compact spot at the centre of the elongated jet, and E is the semi-arched structure extending toward the east (Fig \ref{fig:ModelJets}).

We measured the density of the shocked ISM assuming bremsstrahlung emission in these two regions, as suggested by their spectral indices.
We calculated the emissivity of region D using the integrated flux density within the region and modelled the structure as a cone, with a base matching the beam's dimensions and a height of $\sim$ 12 arcsec. For region E, instead, we used the averaged region flux and the unit volume, following the same approach as for region C in the northern lobe. After converting this value into particle density, we calculated the pre-shock particle density $\rho_0$. 
Similarly to region C in the northern lobe, assuming an electron temperature between $10^4-10^6$ K, we obtained a post-shock particle density $n_e$ in the range $\sim$ 55 -- 150 cm$^{-3}$, resulting in a pre-shock density of $\sim$ 10 -- 40 cm$^{-3}$. This is again consistent with the typical HII and ISM regions electron densities. \cite{Tetarenko_2020} estimated an higher electron density of $n_e \sim 10^4$ cm$^{-3}$: this is not in contrast with our results, as the molecular lines detected in ALMA data trace colder, denser gas within the southern lobe, while our observations reveal the emission from hotter, less dense gas in the same region.

\subsection{Jet age}
\label{sec:jet_age}

We estimated the northern and southern jets velocity and age using the formalism proposed by \cite{Kaiser2004} (see also \citealt{Motta_2025}, \citealt{Atri_2025}), which demonstrates that the velocity of a jet travelling through the surrounding medium is $\sim$ equal to the expansion velocity of the bow-shock formed by the jet-ISM impact. 

We measure a separation of 1.2 arcmin and 1.6 arcmin between the core and the tip of the northern and southern bow-shock regions, respectively, and we obtained a corresponding jet lengths of 3.36$\pm$0.07 pc and 4.41$\pm$0.09 pc, accounting for the binary system's distance and jet inclination angle. We then estimated the two jets ages, which are proportional to the ratio between their lengths and velocities (eq. \ref{eq:jet_age}). We obtained $t_N \in 6.4-26$ kyrs and $t_S \in 11-49$ kyrs, for the northern  and southern lobe, respectively.

The age difference between the northern and southern structure is not surprising: the density is higher toward the north-west direction \citep[see][]{Marti2017nature}, causing the northern jet to interact earlier with the ISM than the southern one. Unsurprisingly, our estimates are consistent with the $\sim 10^4-10^5$ years jet age inferred by \cite{Marti2017nature} using the \cite{Kaiser2004} approach based on VLA observation of GRS\,1758-258.

\subsection{Jet calorimetry}
Following the model from \cite{Kaiser2004}, we calculated the time-averaged jet power $Q_{jet}$ using the quantities derived in the previous sections: $Q_{jet}$ depends on the pre-shock ambient density $\rho_0$, with the jet length $L_{jet}$, and the jet advance velocity $\dot{L}_{jet}$ (see eq. \ref{eq:Q0}). 
We considered the bremsstrahlung emitting region C, D and E, as this assumption allowed us to constrain the pre-shock ISM density.  

We obtained $Q_{jet} \sim 4.4 \times 10^{33}-6.8 \times 10^{35}$ erg s$^{-1}$ for the northern lobe and $Q_{jet} \sim 2.1 \times 10^{34}-3.3 \times 10^{36}$ erg s$^{-1}$ for the southern one. $Q_{jet}$ is very sensitive to the jet opening angle, and hence the two orders of magnitude variation in our estimates partly arise from assuming an opening angle range (1–10 deg), also adopted in \cite{Motta_2025}. The lower limit is set based on the assumption that an excessively small opening angle would inhibit the jet’s expansion. The upper limit is consistent with the opening angles we estimated by measuring the lengths and widths of the structures formed by their interaction with the ISM, $\sim$ 9 deg for the northern jet and $\sim$ 6 deg for the southern jet.

From the inferred jet power, one can also estimate the pressure inside the lobe inflated by the jet (eq. \ref{eq:plobe}). In our case $p_{lobe} \sim 2.1 \times 10^{-12}-2.8 \times 10^{-10}$ erg cm$^{-3}$, where the large range of values arises from the dependence on $Q_{jet}$. These results for $p_{lobe}$ are consistent with the pressure we found in region A and B assuming equipartition, where $p_{B_{eq}} \sim 10^{-10}$ erg cm$^{-3}$.  

For comparison, we estimated the jet power and lobe pressure in region C of the northern lobe, assuming it propagates at the same velocity as region A, as derived from its measured proper motion. Under this assumption, we obtain a jet power of $Q_{jet} \sim 3.1 \times 10^{39}$ to $6.4 \times 10^{40}$ erg s$^{-1}$ and a lobe pressure of $p_{lobe} \sim 2.1 \times 10^{-8}$ to $3.3 \times 10^{-7}$ erg cm$^{-3}$. These significantly larger values arise from the significantly higher velocity of region A with respect to the velocity we considered earlier, which implies taht the shock-heated plasma reached temperatures of $T \sim 10^8$–$10^9$ K under the assumption of bremsstrahlung emission. In turn, such high temperatures imply a higher post-shock electron density, i.e. $n_e \sim$ 300–400 cm$^{-3}$. For consistency with the \cite{Kaiser2004} model, we adopted the velocity inferred from the bremsstrahlung temperature. Additionally, as argued in section \ref{sec5:jet_properties}, region A and C likely result from different episodes of jet activity, thus the velocity of region A does not necessarily reflect that of the jet currently interacting with region C.

\begin{table*}[ht!]
    \centering
    \begin{tabular}{cccc}
    \toprule
       & region C & region D & region E \\
       \midrule
       $n_e$ [cm$^{-3}$] & 55 -- 95 & 80 -- 150 & 55 -- 115 \\
       $Q_{\text{jet}}$ [erg s$^{-1}$] & $4.4 \times 10^{33} - 6.8 \times 10^{35}$ & $ 2.1 \times 10^{34} - 3.3 \times 10^{36}$ & $ 1.3 \times 10^{34} - 2.3 \times 10^{36}$ \\
       $p_{\text{lobe}}$ [erg cm$^{-3}$] & $1.9 \times 10^{-12} - 8.7 \times 10^{-11}$ & $2.9 \times 10^{-12} - 1.4 \times 10^{-10}$ & $2.0 \times 10^{-12} - 9.6 \times 10^{-11}$ \\
    \bottomrule
    \end{tabular}
    \caption{Electron density, time-averaged jet power and lobe pressure inferred for the bremsstrahlung emitting regions, under the assumption of the self-similar model by \cite{Kaiser2004}.}
    \label{tab:regCDE}
\end{table*}


\section{Discussion}\label{sec:discussion}

We observed the position of GRS\,1758-258 for $\sim 7$ hrs with the MeerKAT telescope, and we generated the deepest  L-band radio map of the field to date. The radio image we generated reveals two slightly asymmetric structures, each extending from the compact radio core at a distance of approximately one parsec on either side of the binary.
Such extended structures were already identified as signatures of jet-ISM impacting regions by \cite{Marti2015} using VLA data. Later \cite{Tetarenko_2020} detected molecular emission, revealing the presence of shocked material co-spatial with the southern lobe. Modelling the extended lobe structures around GRS\,1758-258 with the self-similar jet model proposed by \cite{Kaiser2004}, we investigated the ISM properties studying the effects of the jets on the surrounding medium, and estimated the jets' age and the energy transferred from the jet to the ISM.

\subsection{Morphology of the large scale jet structure}
\label{sec5:morph_large_scale}

GRS\,1758-258 is a unique case among the Galactic XRBs with large-scale jets: both jets interact with the ISM inducing the formation of structures at either side of the core, allowing us to probe their properties and to study two distinct regions of the ISM.
The structures on either side differ in both morphology and emission. The northern lobe exhibits a more complex structure, with both synchrotron and bremsstrahlung emitting regions, while the southern lobe is characterised only by thermal bremsstrahlung radiation. We found a slightly higher particle density in the southern regions, and at a greater distance form the core position. This is consistent with the hypothesis proposed by \cite{Marti2017nature}, where they suggest the presence of an ISM density gradient other than the one to the north-south suggested by the association of the CO cloud identified in the \cite{Dame_COsurvey} survey. 

The gas surrounding GRS\,1758-258 is likely multiphase. This could explain the relatively low particle density derived for regions A and B assuming synchrotron emission and equipartition, especially when compared to the higher densities found in bremsstrahlung emitting regions. In our estimates we have adopted a filling factor which accounts for the volume actually occupied by synchrotron emitting gas.   
However, we lack detailed knowledge of the gas distribution in the region, which might be stratified or organized into bubbles of hot gas within cold gas, or vice-versa.
In this scenario it is not surprising that we detect thermal bremsstrahlung emission from the outer edge of the northern lobe - the bow shock structure named region C - which differs in emission properties from the inner regions. The hypothesis of multi-phase gas in the southern lobe is supported by the detection of molecular lines by \cite{Tetarenko_2020}; a similar situation was also observed in the case of GRS\,1915+105, where multi-phase gas could explain the properties of the emission region at the jet termination point \citep{Motta_2025}. The presence of synchrotron emission in the southern lobe cannot be ruled out, given the large uncertainties in our spectral index measurement. \cite{Hardcastle_2005} for instance reported synchrotron emission from both the north and south lobes. While this apparent discrepancy with our findings may be due to intrinsic changes in the emission, the large uncertainties affecting the data prevent us to draw any firm conclusion.

The combined analysis of archival VLA data and new MeerKAT data revealed a complex structure evolution. Some regions (e.g. region B, C, D, E) within the jet structures appear to be static, i.e. likely evolving on time-scales longer than the $\sim$ 20 years sampled by the existing data\footnote{We caution the reader that, due to the faintness of the structures, it is not possible to determine whether small variations are intrinsic to the source or the result of statistical fluctuations couple with the low signal-to-noise ratio of the data. Differences in instrumental configurations may also slightly alter the appearance of the structures.}. Other regions, and in particular region A, corresponding to the jet termination point, exhibit significant proper motion, which corresponds to a de-projected velocity of several thousands of km/s. This is notable as neither in Cyg X-1 nor GRS 1915+105 - the other two examples of galactic XRBs featuring large-scale jet-induced structures - such a short time-scale change has been observed.

\subsection{Jet properties}
\label{sec5:jet_properties}

Table \ref{tab:spect_index} presents a comparison between the flux measurements from VLA \citep{Marti2002} and MeerKAT. In \cite{Marti2002} single flux measurements are reported for the northern and southern structures, whereas in our analysis we provide the integrated flux for the individual regions identified in the two lobes. When adding the contributions for the three regions in the northern lobe and the two regions in the southern lobe, we find that the northern structure is brighter than the southern one.
When measuring the jets' ages, we found an age difference between the northern and the southern jet of $\sim 10^3-10^4$ yrs. \cite{Marti2017nature} ruled out alternative explanations for the difference in distance and brightness
between the northern and southern lobes, concluding that a north-south ISM density gradient is the most plausible scenario. Thus, despite being launch most likely at the same time, the northern and southern jets at first sight may have different ages: the northern jet may 
have interacted with the ISM earlier, and hence travelled a shorter distance, than the southern one. However, the age ranges we estimate broadly overlap, thus we cannot determine with certainty if a difference in age really exist.

Based on the self-similar model, the energy transfer is slightly larger to the south. It is important to note that the model does not estimate the intrinsic jet power - which is expected to be the same for the two 
jets - but rather the energy transferred to the surrounding medium, which depends on the local properties of the ISM. Moreover, we must consider that we are measuring only a fraction of the total jet power: the energy is likely distributed among several processes, including not only the heating of shocked ISM regions but also particle acceleration and the generation of turbulence and bulk motions within the cocoon.
The intrinsic jet power can be in principle inferred modelling the ejecta at early stages, soon after launch when the jet proper motion and deceleration can be directly measured, while the ejecta are still propagating in a low-density environment and may go through deceleration (\citealt{Carotenuto2024}, \citealt{Cooper2025}). The values of the jet energy transferred to the environment that we obtain are a few orders of magnitude lower than the jet powers estimated from the small-scale jets from XRBs. This suggests, on one hand, that the efficiency of energy transfer to the ISM is highly dependent on local environmental conditions, and on the other, that a significant fraction of the jet’s energy is dissipated through non-radiative processes.

The proper motion of region A, combined with the presence of synchrotron emission, suggest that the jet might be currently active in the northern lobe, injecting fresh electrons in the region. One possible interpretation is that the jet has already reached the ISM over-density in the northern lobe, making the synchrotron contribution from the jet detectable. In contrast, in the southern lobe the emitting region is farther away, and the corresponding opposite jet may not yet have become detectable yet. 

The lobe structure surrounding GRS\,1758–258, and particularly the northern lobe, which shows a brighter and more complex structures, is reminiscent of those seen in some AGN jets. Following this analogy, region A would correspond to the hotspot, while region C outlines the cocoon — its outer edge marking the location of the bow-shock. Within this structure, region B may be tracing a backflow (see e.g. \citealt{Matthews_2019}). The formation of a backflow requires a significant density contrast between the jet and the surrounding medium (i.e. $\rho_{jet} < \rho_{ISM}$). We assumed conservation of momentum per unit-mass between the jet and the ISM and the velocity of the proper motion of region A as the velocity of the shocked ISM. We estimated the jet velocity using the distribution of Lorentz factors typical of X-ray binaries \cite[see e.g.][]{Fender2003}. From this, we derived a density ratio $\rho_{jet}/\rho_{ISM} \sim 10^{-4}$. This value becomes even lower when using the post-shock velocity inferred from the bremsstrahlung temperature, further supporting the plausibility of the formation of a  backflow. Simulations of XRBs jets also show that backflows are expected even when the transient jet propagates through a low-density environment near the launch site \citep{Savard_2025}.

Comparing the velocity derived from the proper motion of region A with the thermal velocity assumed for bremsstrahlung emission, we find that the two values are relatively comparable given the large uncertainties involved, although the former is higher. This might indicate that different parts of the lobe are moving and expanding at slightly different velocities, or that the jet is piercing through a pre-existing structure, propagating into a cavity carved out by a previous event (\citealt{Heinz2002}, \citealt{Carotenuto2022}). From the linear fit showed in Fig. \ref{fig:regA_separation} it is possible to infer a rough launching time for the jet terminating in region A, allowing us to derive its age. 
We obtained a jet age in the range of 400-650 years. This value is lower than the age derived from the self-similar model (section \ref{sec:jet_age}) and may support the idea that region A is the result of a more recent jet impact than the bow-shock structure observed in region C, which may have formed due to previous events. 
 

\subsection{Comparison with Cygnus\,X-1 and GRS\,1915+105}

When comparing GRS\,1758-258 to Cygnus\,X-1 and GRS\,1915+105 - the other two BH XRBs exhibiting jet-ISM interaction emission - GRS\,1758-258 shows notable morphological similarities to both sources. In particular, while the bow-shock structure formed by the ISM gas compressed by the approaching jet has been identified in Cygnus\,X-1 and GRS\,1915+105, in the case of GRS\,1758-258, we identify such a structure for both the approaching and receding jets. Moreover, while the overall structure of the northern lobe resembles what is observed in GRS\,1915+105 — featuring a brighter, approximately spherical spot followed by a roughly cylindrical tail within the bow shock feature — the southern lobe appears more similar to the jet-ISM impact region seen near Cygnus\,X-1. This further confirms that the outcome of the jet-ISM interaction is dependent on the properties of the surrounding medium, as, in the case of GRS\,1758-258, an ISM density gradient has been suggested to explain the structural differences between the two lobes (see section \ref{sec:jet_age}). 

Despite the morphological similarities, the lobes observed around GRS\,1758-258 are closer and less extended in size that the structures identified in GRS\,1915+105 and Cygnus\,X-1. Similarly the ages of the structures around GRS\,1915+105 and Cygnus\,X-1 are significantly higher than in GRS\,1758-258 (tens to hundreds of kilo-years). This may suggest that GRS\,1915+105 and Cygnus\,X-1 have remained active for longer, allowing the jets sufficient time to carve out cavities and form extend bow-shock structures, or that repeated jet events have emptied the surrounding of these systems, allowing for the most recent events to induce the formation of structures much further away. It seems unlikely that the difference in the jet structures between the different sources could be ascribed to the local environment, since the particle density of the ISM surrounding GRS\,1758-258 is similar to the particle density inferred for both GRS\,1915+105 and Cygnus\,X-1, and to the typical ISM densities near the Galactic Plane. 

The detection of proper motion in a synchrotron emitting structure provides further evidence for the presence of an active jet moving away from the launching site. In GRS\,1915+105, synchrotron emission is observed in certain regions of the jet-ISM impact structure, although no detectable proper motion has been detected, possibly due to the limited angular resolution possible with MeerKAT and the large distance to the source. Cygnus\,X-1 shows neither proper motion nor synchrotron emission, but only bremsstrahlung radiation likely produced by gas that was heated during a past jet-ISM interaction, which may indicate that the structure we observe has not been significantly energised over the past several thousand of years. However, a weak or unresolved synchrotron contribution cannot be excluded, leaving open the possibility that some level of ongoing or recent energisation may still be occurring.

When considering the inferred energetics of the lobes, we obtain a time-averaged jet power consistent with the estimate for Cygnus\,X-1 but lower than that of GRS\,1915+105. GRS\,1915+105 is a peculiar BH-XRBs, known for having the largest accretion disc among such systems and for accreting at a rate close to the Eddington limit when active \citep{Done2004}. This could explain why it exhibits more energetic outflows compared to the other two sources. It is also important to consider that the jet power estimated using the self-similar model depends on the electron density of the ISM region interacting with the jet, as discussed in section \ref{sec5:jet_properties}. In the case of GRS\,1915+105, the jet impacts a region with a density of hundreds of particles cm$^{-3}$, which likely results in a more efficient energy transfer.


\section{Summary and Conclusions}\label{sec: conclusions}

MeerKAT L-band observation of GRS\,1758-258 reveal extended lobes structures along the direction of the bipolar jet of the BH binary. These features originate from the impact between the jet launched from the accreting binary and the surrounding ISM. Jet-ISM impact sites serves as excellent laboratories for investigating the accretion feedback and the accretion-ejection coupling in black hole accreting systems. GRS\,1758-258 represents a rather unique case, as both the approaching and receding jets are observed to interact with the ISM at large distances from the launching site. The only other Galactic example is the BH-XRB 1E 1740.7$-$2942, which exhibits a morphology reminiscent of that of FRI AGN sources (see e.g. \citealt{Fanaroff1974}), whereas GRS 1758$-$258 is more reminiscent of an FRII source. 

Thanks to the high sensitivity of the MeerKAT observations, we were able to resolve different emitting regions in both the northern and the southern lobes. Applying the self-similar model by \cite{Kaiser2004}, we constrained the jet power transferred to the surrounding ISM. The presence of jet-ISM interaction structures on both the jet and the counter jet sides allowed us to infer different ISM properties around GRS\,1758-258. Furthermore, a comparison between our estimates of the jet energy transferred to the environment and the average jet power estimated from the small-scale transient jets from XRBs, suggests that only a fraction of the jet's energy is effectively deposited into the ISM, where this fraction depends on the local ISM conditions.

GRS\,1758-258 also shows a slightly lower transferred energy compared with the other two notable Galactic sources associated with jet-ISM interaction structures, GRS\,1915+105 and Cygnus\,X-1. The jet-driven features in GRS\,1758-258 are less extended and located closer to the position of the binary. Combined with the inferred ages and the size of the jet-induced structures (smaller in GRS\,1758-258 than in GRS\,1915+105 and Cygnus\,X-1), this suggests that the lobes in GRS\,1758-258 are younger, and the jets have not yet had sufficient time to transfer more energy and/or to carve out large, parsec-scale bow-shock structures in the surrounding medium. Moreover, the detection of synchrotron emission and proper motion in the northern jet hints to the presence of an active jet feeding electrons to the medium, other than thermal electrons heated up by past jet activity. We argue that the structure around GRS\,1758-258 may represent an earlier evolutionary stage of the jet-ISM interaction, compared to what is seen in GRS\,1915+105 and Cygnus\,X-1.


\begin{acknowledgements}

IM and SEM acknowledg support from the INAF Fundamental Research Grant (2022) EJECTA.
PLLE and JM acknowledge support from grant PID2022-136828NB-C42 funded by the Spanish MCIN/AEI/ 10.13039/501100011033 and “ERDF A way of making Europe.

The MeerKAT telescope is operated by the South African Radio Astronomy Observatory, which is a facility of the National Research Foundation, an agency of the Department of Science and Innovation.

We acknowledge the use of the Ilifu cloud computing facility – www.ilifu.ac.za, a partnership between the University of Cape Town, the University of the Western Cape, Stellenbosch University, Sol Plaatje University, and the Cape Peninsula University of Technology. The Ilifu facility is supported by contributions from the Inter-University Institute for Data Intensive Astronomy (IDIA – a partnership between the University of Cape Town, the University of Pretoria and the University of the Western Cape), the Computational Biology division at UCT and the Data Intensive Research Initiative of South Africa (DIRISA).\\

\end{acknowledgements}

\section*{Data Availability}

The un-calibrated MeerKAT visibility data presented in this paper are publicly available in the archive of the South African Radio Astronomy Observatory at \url{https://archive.sarao.ac.za}. 
The python source code used to perform the calculations is available upon reasonable request to the authors. 

%
%

\bibliographystyle{aa.bst}
\bibliography{biblio} 

\newpage
\begin{appendix} 

\section{Additional informations}\label{App:App1}
\subsection{Synchrotron emission at equipartition}
\label{App:synch}
Synchrotron radiation is produced by a relativistic plasma composed of a magnetic field and a population of relativistic electrons: the total energy is thus given by accounting for both the energy of the magnetic field and of the electrons. The condition of minimum energy occurs when these two energies are almost equal, a state referred to as equipartition\footnote{For a complete derivation, see \url{https://github.com/robfender/ThunderBooks}}:
\begin{equation}
    E_B = \dfrac{3}{4} \eta E_e
    \label{eq:equipart}
\end{equation}
here $\eta$ represents the energy content stored in the protons accompanying the electrons. From the \ref{eq:equipart} is possible to derive the magnetic field at the equipartition $B_{eq}$, resolving the dependencies of $E_B$ and $E_e$ on B:
\begin{equation}
    B_{eq} = \left(6\pi\dfrac{\eta}{f}c_{12}\dfrac{L}{V}\right)^{2/7}
    \label{eq:Beq}
\end{equation}
where $f$ represents the filling factor, i.e. the fraction of the physical source's volume occupied by the magnetic field and the electrons, $V$ denotes the volume and $c_{12}$ is a pseudo-constant that contains the informations about the frequency range and the slope of the energy spectrum within this range \citep{Pacho}. For a power law distribution of index $-p$ between energy limits $E_1$ and $E_2$, the total number of electrons is given by:
\begin{equation}
    N = \int^{E_2}_{E_1} N_0E^{-p}dE = \dfrac{N_0}{1-p}\left(E_2^{1-p}-E_1^{1-p}\right)
    \label{eq:N}
\end{equation}
For an electron of energy E in a magnetic field B the rate of energy loss by synchrotron radiation is 
\begin{equation}
    -\dfrac{dE}{dt} = c_2B^2E^2
\end{equation}
here $c_2 = 2.37 \times 10^{-3}$ (cgs units). Therefore the luminosity observed from $N(E) = E^{-p}$ in a field B is given by:
\begin{multline}
    L = - \int^{E_2}_{E_1} \dfrac{dE}{dt}N(E)dE = \\
    = N_0c_2B^2\int^{E_2}_{E_1} E^{2-p}dE = \dfrac{N_0c_2B^2}{3-p}\left(E_2^{3-p}-E_1^{3-p}\right)
    \label{eq:LN0}
\end{multline}
Writing the electric field $E$ as a function of frequency we obtain $E = \dfrac{\nu^{1/2}}{c_1^{1/2}B^{1/2}}$, with the constant $c_1$ equal to $6.27\times10^{18}$ in cgs units. In the convention we are using, the spectral index is defined as $\alpha = \dfrac{1-p}{2}$, assuming a power-law dependence for the monochromatic flux density $S_{\nu} \propto \nu^{\alpha}$. Solving for $N_0$ in \ref{eq:LN0} and replacing it in the \ref{eq:N}, after converting energies into frequencies:
\begin{equation}
    N = \dfrac{(3-p)}{(1-p)}\dfrac{c_1L}{c_2B}\dfrac{\left[\nu_2^{(1-p)/2}-\nu_1^{(1-p)/2}\right]}{\left[\nu_2^{(3-p)/2}-\nu_1^{(3-p)/2}\right]}
    \label{eq:Nsyn}
\end{equation}
The total pressure inside the volume V for minimum energy conditions is (\cite{Kaiser2004}):
\begin{equation}
    p_{E_{min}} = \dfrac{7}{9} \dfrac{B_{eq}^2}{8\pi}(k+1)
    \label{eq:pEmin}
\end{equation}
where $k$ is the internal energy stored in the particles not contributing to synchrotron emission to the sum of the energies $E_B + E_e$.
\subsection{Bremsstrahlung emission}
In the case of thermal bremsstrahlung emission from ionised hydrogen, the monochromatic emissivity is (\cite{Longair1994}):
\begin{equation}
    \epsilon_{\nu} = \dfrac{L_{\nu}}{V} = C_{\text{radio}}g(\nu,T)\dfrac{n_e^2}{\sqrt{T}}\exp\left(\dfrac{h\nu}{k_bT}\right) \ \dfrac{\text{erg}}{\text{s\ cm}^3\text{Hz}}
    \label{eq:epsnu}
\end{equation}
with $C_{\text{radio}} = 6.8 \times 10^{-38}$ erg \ s$^{-1}$ cm$^{-3}$ Hz$^{-1}$ and the Gaunt factor:
\begin{equation}
    g(\nu,T) = \dfrac{\sqrt{3}}{2\pi}\left[\ln\left(\dfrac{128\epsilon_0k_b^3T^3}{m_ee^4\nu^2Z^2}\right)-\gamma^{1/2}\right]
\end{equation}
where we used $\epsilon_0=1$ and $Z=1$ and $\gamma$ is the Euler's constant. The expression for the electron density number is then:
\begin{equation}
    n_e = \sqrt{\dfrac{\epsilon_{\nu}}{g(\nu,T)\sqrt{T}C_{\text{radio}}exp\left(\frac{h\nu}{k_bT}\right)}}
    \label{eq:ne}
\end{equation}

\subsection{Jet age and calorimetry}
\cite{Kaiser2004} showed that for an ISM with a constant density the jet length evolves a function of the time t:
\begin{equation}
    L_{\text{jet}} = C_1 \left(\dfrac{Q_0}{\rho_0}\right)^{1/5}t^{3/5}
    \label{eq:Lj}
\end{equation}
where $C_1$ is a dimensionless constant that depends on the thermodynamic properties of the jet and the aspect ratio of the lobe inflated by it, defined as the ratio between the lobe's length and width (assuming a cylindrical shape): 
\begin{equation}
    C_1 = \left(\dfrac{C_2}{C_3\theta^2}\dfrac{(\Gamma_x+1)(\Gamma_c-1)(5-\beta)^3}{18\left[9\left\{\Gamma_c+(\Gamma_c-1)\frac{C_2}{4\theta^2}\right\}-4-\beta\right]}\right)
\end{equation}
where $\Gamma_x$, $\Gamma_c$ are the adiabatic indices of the material in the surrounding ISM and in the lobe cavity, respectively, $\theta$ denotes the jet opening angle, $\beta$ the density profile index. The terms $C_2$ and $C_3$ are defined as:
\begin{equation}
    C_2 = \left(\dfrac{(\Gamma_c-1)(\Gamma_{jet}-1)}{4\Gamma_c}+1\right)^{\frac{\Gamma_c}{\Gamma_c-1}}\dfrac{(\Gamma_{jet}+1)}{(\Gamma_{jet}-1)}
\end{equation}
\begin{equation}
    C_3 = \dfrac{\pi}{4R_{ax}^2}
\end{equation}
with $\Gamma_{jet}$ = adiabatic index of the material within the jet, $R_{ax}=\sqrt{\dfrac{1}{4}\dfrac{C_2}{\theta^2}}$ = aspect ratio of the lobe. In our case, we assumed $\beta=0$, indicating a uniform density in both the jet and in the external medium, and set $\Gamma_x$ = $\Gamma_c$ = $\Gamma_{jet} = 5/3$. \\
Deriving and solving for the time equation \ref{eq:Lj} we obtain:
\begin{equation}
    t = \dfrac{3L_{\text{jet}}}{5\dot{L}_{\text{jet}}}
    \label{eq:jet_age}
\end{equation}
from which is possible to estimate the jet age. We determined the velocity $\dot{L}_{jet}$ considering its dependence on temperature in the case of a strongly shocked monoatomic gas:
\begin{equation}
    \dot{L} = \sqrt{\dfrac{16k_b}{3m_p}T}
    \label{eq:Ldot}
\end{equation}
where $m_p$ = proton mass. We setted the temperature range under the assumption of bremsstrahlung emission for the bow-shock structures. Finally, using the \ref{eq:Lj} and \ref{eq:jet_age} we derive the following expression for the time-averaged jet power:
\begin{equation}
    Q_0 = \left(\dfrac{5}{3}\right)^3\dfrac{\rho_0}{C_1^5}L_{\text{jet}}^2\dot{L_{\text{jet}}}^3
    \label{eq:Q0}
\end{equation}
The pressure inside the lobe is given by:
\begin{equation}
    p_{\text{lobe}} = 0.0675\dfrac{C_1^{10/3}}{R_{ax}^2}\left(\dfrac{\rho_0Q_0^2}{L_{\text{jet}}^4}\right)^{1/3}
    \label{eq:plobe}
\end{equation}

\subsection{Jet velocity}

\begin{figure}[ht!]
    \centering
    \includegraphics[width=\columnwidth]{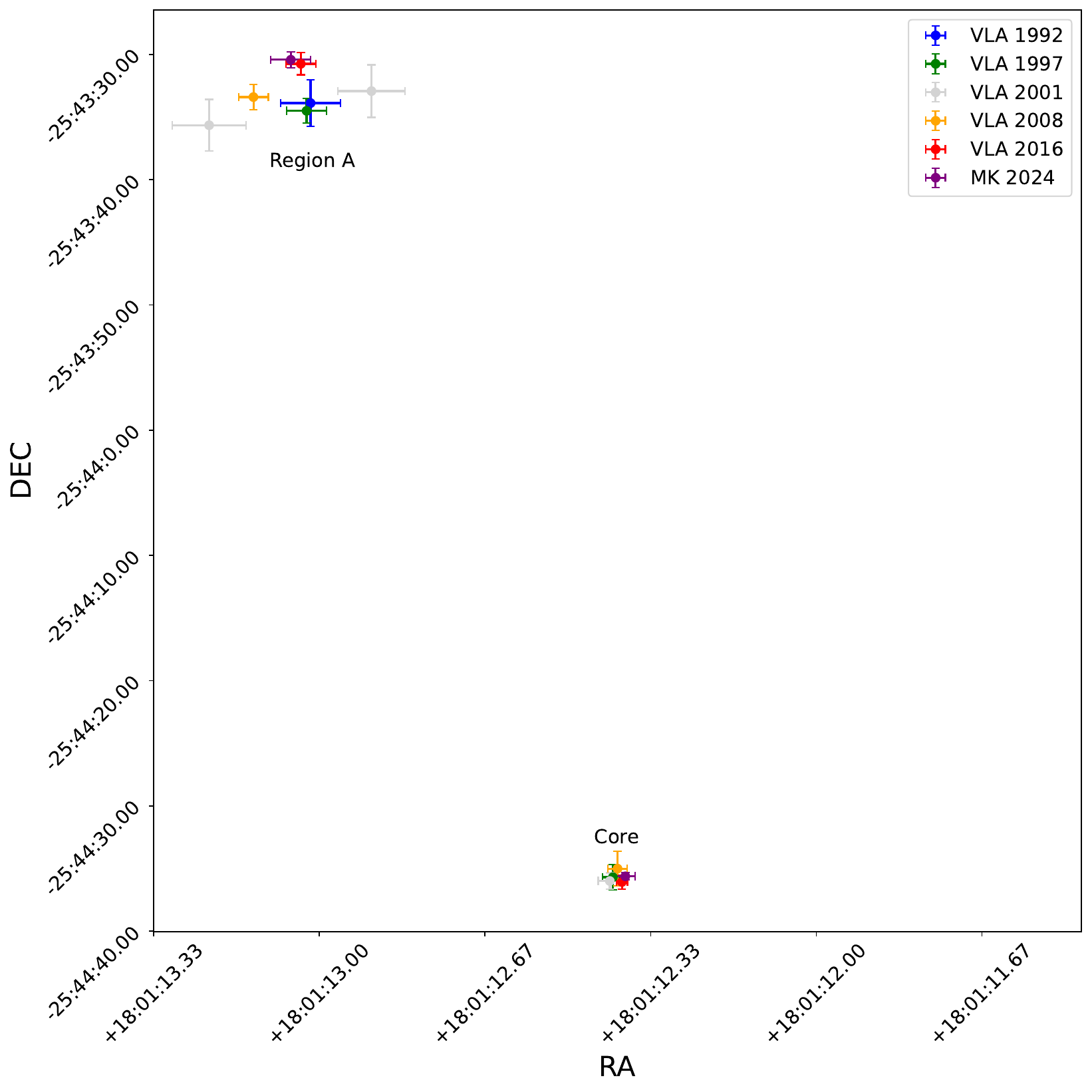}
    \caption{Core (bottom centre) and bright northern spot (region A, top left) across different epochs. Unlike the core, the northern bright spot exhibits a clear proper motion towards the $\sim$ north-east direction.}
    \label{figA:regA-core-pos}
\end{figure}

\subsection{Additional tables}

\begin{table}[ht!]
    \centering
    \begin{tabular}{cccc}
    \toprule
     Regions/Flux density & S$_{1.18GHz}$ [mJy] & S$_{1.39GHz}$ [mJy] & S$_{1.61GHz}$ [mJy]\\
     \midrule
     Core & 0.43$\pm$0.03 & 0.44$\pm$0.02 & 0.41$\pm$0.02 \\
     \rowcolor{gray!30}
     \multicolumn{4}{c}{Northern lobe} \\
     A & 0.35$\pm$0.03 & 0.28$\pm$0.03 & 0.28$\pm$0.03 \\
     B & 0.37$\pm$0.04 & 0.31$\pm$0.03 & 0.30$\pm$0.03\\
     C & 0.52$\pm$0.05 & 0.36$\pm$0.04 & 0.43$\pm$0.04 \\
     \rowcolor{gray!30}
     \multicolumn{4}{c}{Southern lobe} \\
     D & 0.14$\pm$0.01 & 0.13$\pm$0.01 & 0.13$\pm$0.01 \\
     E & 0.53$\pm$0.05 & 0.44$\pm$0.04 & 0.54$\pm$0.05 \\
     \bottomrule
    \end{tabular}
    \caption{In-band fluxes for each region used to calculate their spectral indices.}
    \label{tab:flux_spectra}
\end{table}

\begin{table*}[ht!]
    \centering
    \begin{tabular}{ccc}
    \toprule
     year & core [RA, DEC] & region A [RA,DEC] \\
     \rowcolor{gray!30}
     1992 & / & 18$^h$01$^m$13.02 $\pm$ 0.03$^s$, -25$^{\circ}$43$'$34 $\pm$ 2$''$\\
     1997 & 18$^h$01$^m$12.41 $\pm$ 0.02$^s$, -25$^{\circ}$44$'$35 $\pm$ 1$''$ & 18$^h$01$^m$13.03 $\pm$ 0.03$^s$, -25$^{\circ}$43$'$34 $\pm$ 1$''$  \\
     \rowcolor{gray!30}
          & & 18$^h$01$^m$13.22 $\pm$ 0.07$^s$, -25$^{\circ}$43$'$36 $\pm$ 2$''$\\
    \rowcolor{gray!30}
     2001 & 18$^h$01$^m$12.41 $\pm$ 0.02$^s$, -25$^{\circ}$44$'$36.0 $\pm$ 0.6$''$ & \\
     \rowcolor{gray!30}
          & & 18$^h$01$^m$12.89 $\pm$ 0.07$^s$, -25$^{\circ}$43$'$33 $\pm$ 2$''$  \\
     2008 & 18$^h$01$^m$12.40 $\pm$ 0.02$^s$, -25$^{\circ}$44$'$35 $\pm$ 1$''$ & 18$^h$01$^m$13.12 $\pm$ 0.01$^s$, -25$^{\circ}$43$'$34 $\pm$ 1$''$   \\
     \rowcolor{gray!30}
     2016 & 18$^h$01$^m$12.39 $\pm$ 0.01$^s$, -25$^{\circ}$44$'$36.1 $\pm$ 0.5$''$ & 18$^h$01$^m$13.04 $\pm$ 0.02$^s$, -25$^{\circ}$43$'$30.7 $\pm$ 0.9$''$\\
     2024 & 18$^h$01$^m$12.38 $\pm$ 0.02$^s$, -25$^{\circ}$44$'$35.6 $\pm$ 0.3$''$ & 18$^h$01$^m$13.06 $\pm$ 0.04$^s$, -25$^{\circ}$43$'$30.4 $\pm$ 0.6$''$ \\
     \bottomrule
    \end{tabular}
    \caption{Fitted positions of the core and the northern bright spot (region A) across the different epochs.}
    \label{tab:regA_pos}
\end{table*}

\begin{table*}[ht!]
    \centering
    \begin{tabular}{ccc}
    \toprule
    \multicolumn{3}{c}{\textbf{Source-related parameters}} \\
    \toprule
     & Distance & 8.5 kpc \\
     & Jet inclination angle &  61$\pm$2 deg\\
     & Jet opening angle & 1-10 deg\\
    \toprule
    \multicolumn{3}{c}{\textbf{Northern jet}} \\
    \rowcolor{gray!30}
    \multicolumn{3}{c}{Region A - non thermal emission region} \\
    \multirow{4}{*}{Measured}
    & Radius & $\sim$ 6.3 arcsec $\sim$ 0.1 pc\\
    & Volume (sphere) & $\sim 2.1\times 10^{54}$ cm$^3$\\
    & Flux density & $0.31\pm0.03$ mJy \\
    & Integrated luminosity & $3.6^{+1.5}_{-0.5} \times 10^{29}$ erg s$^{-1}$ \\
    \hdashline
    \multirow{3}{*}{Inferred (equipartition)}
    & Magnetic density & $\gtrsim 4.5 \times 10^{-5}$ G\\
    & Pressure & $\gtrsim 6.5 \times 10^{-11}$ erg cm$^{-3}$\\
    & Synchrotron electron density &  $\lesssim 6.9 \times 10^{-7}$ cm$^{-3}$\\
    \rowcolor{gray!30}
    \multicolumn{3}{c}{Region B - non thermal emission region} \\
    \multirow{4}{*}{Measured}
    & Thickness & $\sim$ 3.2 arcsec $\sim$ 0.1 pc\\
    & Length & $\sim$ 24.5 arcsec $\sim$ 1.1 pc \\
    & Volume (cylinder) & $\sim 1.8\times 10^{54}$ cm$^3$\\
    & Flux density & $0.33\pm0.03$ mJy \\
    & Integrated luminosity & $4.5^{+1.8}_{-0.9} \times 10^{29}$ erg s$^{-1}$\\
    \hdashline
    \multirow{3}{*}{Inferred (equipartition)}
    & Magnetic density & $\gtrsim 4.8 \times 10^{-5}$ G\\
    & Pressure & $\gtrsim 7.2 \times 10^{-11}$ erg cm$^{-3}$\\
    & Synchrotron electron density & $\lesssim 5.8 \times 10^{-7}$ cm$^{-3}$ \\
    \rowcolor{gray!30}
    \multicolumn{3}{c}{Region C - bow shock structure} \\ 
    \multirow{4}{*}{Measured}
    & Thickness & $\sim$ 16.2 arcsec $\sim$ 0.7 pc\\
    & Length & $\sim$ 1.2 arcmin $\sim$ 3.4 pc\\
    & Volume  & $\sim 4.3 \times 10^{55}$ cm$^3$\\
    & Averaged flux density & $0.042\pm0.004$ mJy beam$^{-1}$ \\
    \hdashline
    \multirow{5}{*}{Inferred (self-similar model)}
    & Shock-compressed gas electron density & 55 -- 95 cm$^{-3}$\\
    & ISM gas density & 10 -- 25 cm$^{-3}$\\
    & Lobe pressure  & $1.9 \times 10^{-12} - 8.7 \times 10^{-11}$ erg cm$^{-3}$\\
    & Power transferred  & $4.4 \times 10^{33} - 6.8 \times 10^{35}$ erg s$^{-1}$\\
    & Jet age & 6 -- 26 kyrs \\

    \toprule
    \multicolumn{3}{c}{\textbf{Southern jet}} \\
    \rowcolor{gray!30}
    \multicolumn{3}{c}{Region D - thermal emission region} \\
    \multirow{4}{*}{Measured}
    & Radius & $\sim$ 4.8 arcsec $\sim$ 0.2 pc\\
    & Height & $\sim$ 12.3 arcsec $\sim$ 0.6 pc\\
    & Volume & $\sim 6.7 \times 10^{53}$ cm$^3$\\
    & Flux density & $0.14\pm0.01$ mJy\\
    \hdashline
    \multirow{4}{*}{Inferred (self-similar model)}
    & Shock-compressed gas electron density & 80 -- 155 cm$^{-3}$\\
    & ISM gas density & 20 -- 40 cm$^{-3}$\\
    & Lobe pressure  & $2.9 \times 10^{-12} - 1.4 \times 10^{-10}$ erg cm$^{-3}$\\
    & Power transferred  & $2.1 \times 10^{34} - 3.3 \times 10^{36}$ erg s$^{-1}$\\
    \rowcolor{gray!30}
    \multicolumn{3}{c}{Region E - bow shock structure} \\
    \multirow{4}{*}{Measured}
    & Thickness & $\sim$ 13.5 arcsec $\sim$ 0.6 pc\\
    & Length & $\sim$ 1.6 arcmin $\sim$ 4.4 pc\\
    & Volume & $\sim 4.0 \times 10^{55}$ cm$^3$\\
    & Averaged flux density & $0.037\pm0.004$ mJy beam$^{-1}$ \\
    \hdashline
    \multirow{5}{*}{Inferred (self-similar model)}
    & Shock-compressed gas electron density & 55 -- 115 cm$^{-3}$\\
    & ISM gas density & 12 -- 30 cm$^{-3}$\\
    & Lobe pressure  & $2.0 \times 10^{-12} - 9.6 \times 10^{-11}$ erg cm$^{-3}$\\
    & Power transferred  & $1.3 \times 10^{34} - 2.3 \times 10^{36}$ erg s$^{-1}$\\
    & Jet age & 11 -- 46 kyrs \\
    \bottomrule
    \end{tabular}
    \caption{Summary table of the measured and inferred properties of the northern jet, southern jet and surrounding medium.}
    \label{tab:total_table_south}
\end{table*}

\end{appendix}

\end{document}